\documentclass[apjl]{emulateapj}

\newcommand{\he}{HE\,0450--2958}

\shorttitle{The QSO HE\,0450--2958}
\shortauthors{K.\ Jahnke et al.}

\begin{document}

\title{The QSO HE\,0450--2958: Scantily dressed or heavily robed?\\ A
  normal quasar as part of an unusual ULIRG.}

\author{Knud Jahnke\altaffilmark{1}, David Elbaz\altaffilmark{2}, Eric
  Pantin\altaffilmark{2}, Asmus B\"ohm\altaffilmark{3}$^,$\altaffilmark{4}, Lutz
  Wisotzki\altaffilmark{3}, Geraldine Letawe\altaffilmark{5}, Virginie
  Chantry\altaffilmark{5}, Pierre-Olivier Lagage\altaffilmark{2}}

\email{jahnke@mpia.de}

\altaffiltext{1}{Max-Planck-Institut f\"ur Astronomie,
K\"onigstuhl 17, D-69117 Heidelberg, Germany}
\altaffiltext{2}{CEA Saclay/Service d'Astrophysique, Laboratoire AIM,
CEA/DSM/IRFU-CNRS-Universit\'e Paris Diderot, F-91191 Gif-sur-Yvette
C\'edex, France}
\altaffiltext{3}{Astrophysikalisches Institut Potsdam, An der Sternwarte
16, D-14482 Potsdam, Germany} 
\altaffiltext{4}{Institut f\"ur Astro- und Teilchenphysik, Universit\"at
  Innsbruck, Technikerstra{\ss}e 25/8,  A-6020 Innsbruck, Austria}
\altaffiltext{5}{Institut d'Astrophysique et G\'eophysique, Universit\'e
de Li\`ege, All\'ee du 6 Ao\^ut, 17 Sart Tilman (Bat. B5C), B-4000
Li\`ege, Belgium}

\begin{abstract}
  The luminous $z=0.286$ quasar \he\ is interacting with a companion galaxy at
  6.5~kpc distance and the whole system radiates in the infrared at the level
  of an ultraluminous infrared galaxy (ULIRG). A so far undetected host galaxy
  triggered the hypothesis of a mostly ``naked'' black hole (BH) ejected from
  the companion by three-body interaction. We present new HST/NICMOS 1.6$\mu$m
  imaging data at 0\farcs1 resolution and VLT/VISIR 11.3$\mu$m images at
  0\farcs35 resolution that are for the first time resolving the system in the
  near- and mid-infrared. We combine these data with existing optical HST and
  CO maps.
(i) At 1.6$\mu$m we find an extension N-E of the quasar nucleus that is likely
  a part of the host galaxy, though not its main body. If true, a combination
  with upper limits on a main body co-centered with the quasar brackets the
  host galaxy luminosity to within a factor of $\sim$4 and places \he\
  directly onto the $M_\mathrm{BH}-M_\mathrm{bulge}$-relation for nearby
  galaxies.
(ii) A dust-free line of sight to the quasar suggests a low dust
  obscuration of the host galaxy, but the formal upper limit for star
  formation lies at 60~M$_\odot$/yr. \he\ is consistent with lying at the
  high-luminosity end of Narrow-Line Seyfert 1 Galaxies, and more exotic
  explanations like a ``naked quasar'' are unlikely.
(iii) All 11.3$\mu$m radiation in the system is emitted by the quasar
  nucleus. It has warm ULIRG-strength IR emission powered by black hole
  accretion and is radiating at super-Eddington rate,
  $L/L_\mathrm{Edd}=6.2^{+3.8}_{-1.8}$, or 12~$M_\odot$/year.
(iv) The companion galaxy is covered in optically thick dust and is not a
  collisional ring galaxy. It emits in the far infrared at ULIRG strength,
  powered by Arp220-like star formation (strong starburst-like). An
  M82-like SED is ruled out.
(v) With its black hole accretion rate \he\ produces not enough new stars
  to maintain its position on the
  $M_\mathrm{BH}-M_\mathrm{bulge}$-relation, and star formation and black
  hole accretion are spatially disjoint. This relation can either only be
  maintained averaging over a longer timescale ($\la$500~Myr) and/or the bulge
  has to grow by redistribution of preexisting stars.
(vi) Systems similar to \he\ with spatially disjoint ULIRG-strength star
  formation and quasar activity might be common at high redshifts but at
  $z<0.43$ we only find $<$4\% (3/77) candidates for a similar
  configuration.
\end{abstract}


\section{Introduction}
In the current framework of galaxy evolution, galaxies and black holes are
intimately coupled in their formation and evolution. The masses of
galactic bulges and their central black holes (BHs) in the local Universe
follow a tight relation \citep[e.g.][]{haer04} with only 0.3~dex
scatter. Currently it is not clear how this relation comes about and if
and how it evolved over the last 13 Gyrs, but basically all semi-analytic
models now include feedback from active galactic nuclei (AGN) as a key
ingredient to acquire consensus with observations
\citep[e.g.][]{hopk06c,some08}. In these models it is assumed that black
hole growth by accretion and energetic re-emission from the ignited AGN
back into the galaxy can form a self regulating feedback chain. This
feedback loop can potentially regulate or possibly also truncate star
formation and in this process create and maintain the red/blue
color--magnitude bimodality of galaxies. In this light, any galaxy with an
abnormal deviation from the $M_\mathrm{BH}$--$M_\mathrm{bulge}$-relation
will be an important laboratory for understanding the coupling mechanisms
of black hole and bulge growth. It will set observational limits for these
models, and constrain the time-lines and required physics involved.

Since the early work by \citet{bahc94,bahc95b} on QSO host galaxies with
the {\em Hubble Space Telescope (HST)} and the subsequently resolved
dispute about putatively ``naked'' QSOs \citep{mcle95a}, no cases for QSOs
without surrounding host galaxies were found -- when detection limits were
correctly interpreted. Only recently the QSO \he\ renewed the discussion,
when \citet{maga05} made a case for a 6$\times$ too faint upper limit of
the host galaxy of \he\ with respect to the
$M_\mathrm{BH}$--$M_\mathrm{bulge}$-relation. In light of a number of
competing explanations for this, the nature of the \he\ system needs to be
settled.
\medskip

The QSO \he\ (a.k.a.\ IRAS~04505--2958) at a redshift of $z=0.286$ was
discovered by \citet{low88} as a warm IRAS source. \he\ is a radio-quiet
quasar, with a distorted companion galaxy at 1\farcs5 (=6.5~kpc) distance
at the same redshift, likely in direct interaction with the QSO
\citep{cana01}. The combined system shows an infrared luminosity of an
ultraluminous infrared galaxy (ULIRG, $L_\mathrm{IR}>10^{12}$~L$_\odot$).

\begin{figure*}
\includegraphics[width=\textwidth]{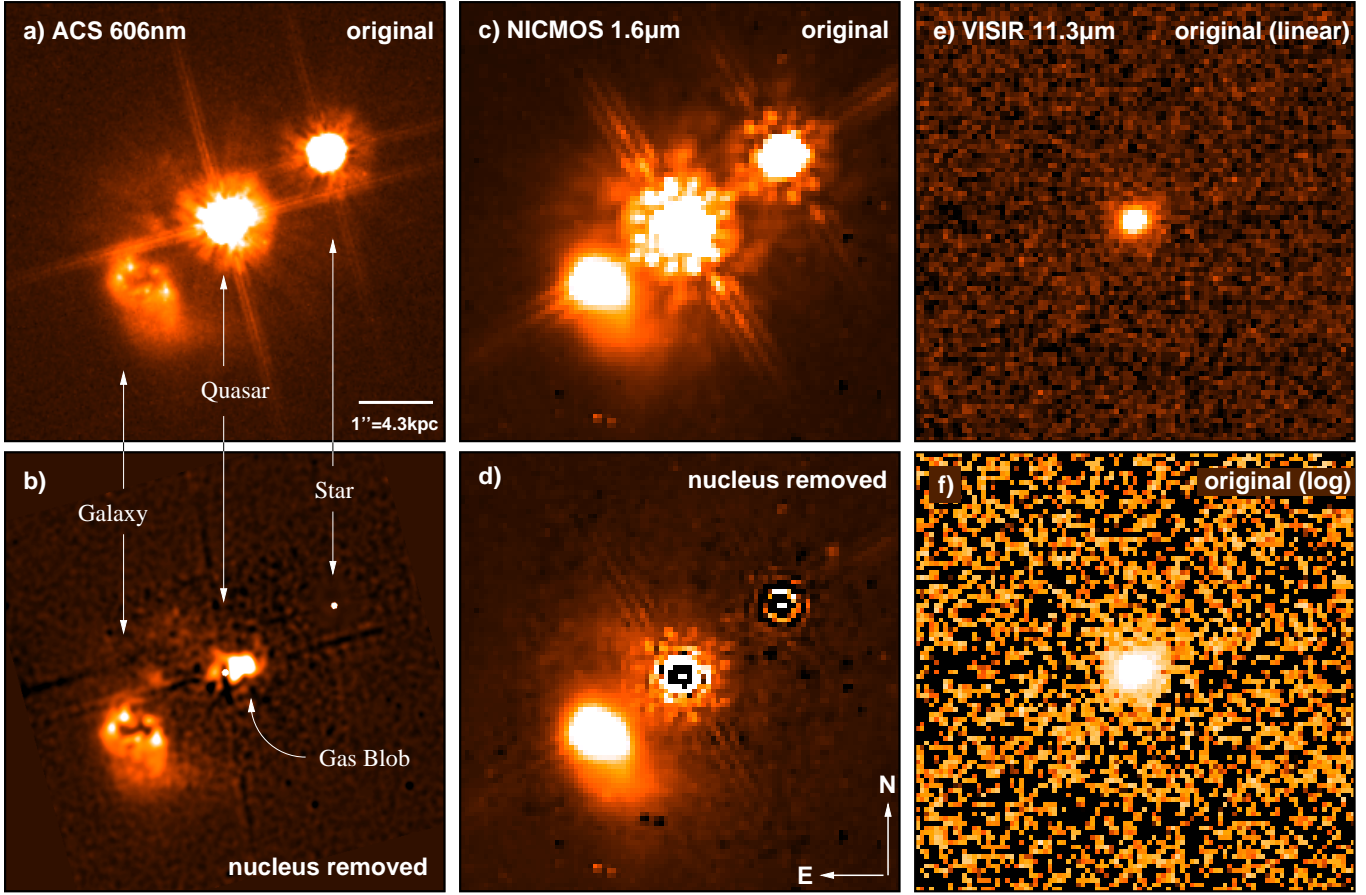}
\caption{\label{fig:allwave} 
The \he\ system as seen in the existing optical images, and our new NIR and
MIR data with the same scale and orientation (field size is 7\arcsec, N is up
and E to the left). Shown are {\em (a)} the original ACS HRC image in the $V$
band (F606W) by \citet{maga05}, {\em (b)} the deconvolved ACS HRC image, {\em
(c)} the original NICMOS $H$ band (F160W), {\em (d)} the NICMOS image after
image decomposition and subtraction of the nuclear component and foreground
star, and the VISIR PAH2 filter image at 11.3$\mu$m {\em (e)} in linear
stretch and {\em (f)} in logarithmic stretch. See Fig.~\ref{fig:dataimages}
for more NICMOS images. The VISIR image {\em (e+f)} is diffraction limited but
shows a single unresolved 62.5~mJy point source, the AGN. No other source in
the field is detectable above a point-source flux density of 3~mJy
(5$\sigma$), the companion galaxy is also undetected.
In the ACS image {\em (b)} the PSF-distributed flux of the point like QSO
nucleus and foreground star have been condensed into the two white points. The
``Blob'' 0\farcs5 from the QSO is made only of emission line light
\citep{leta08} -- and thus not visible in the NIR since we have no similar
line in the F160W bandpass. The companion galaxy 1\farcs5 (=6.5~kpc) to the
south-west has $M_V=-23$ \citep{maga05}. It has a complex structure, but as a
difference to the optical it is clearly peaked towards the center in the
NIR. Towards the SW of the companion the tidal arm described by \citet{cana01}
is visible.
}
\end{figure*}

\he\ was observed with the {Hubble Space Telescope (HST)} and its WFPC2
camera \citep{boyc96} in F702W (=$R$ band) and ACS camera \citep{maga05}
in F606W (=$V$ band), both observations did not allow to detect a host
galaxy centered on the quasar position within their limits
(Figure~\ref{fig:allwave}, left column).  \citet{maga05} estimated an
expected host galaxy brightness if \he\ was a normal QSO system that
obeyed the $M_\mathrm{BH}$--$M_\mathrm{bulge}$-relation in the local
Universe and given a BH mass estimate or luminosity of the QSO. They
concluded that the ACS F606W detection limits were six times fainter than
the expected value for the host galaxy, which qualified \he\ to be very
unusual.


\citet{maga05} sparked a flurry of subsequent papers to explain the
undetected host galaxy to black hole relation. Over time three different
alternative explanations have been put forward and were substantiated:

\begin{enumerate}
  
\item \he\ is a normal QSO nucleus, but with a massive black hole residing
  in an under-massive host galaxy. The system is lying substantially off
  the local $M_\mathrm{BH}$--$M_\mathrm{bulge}$-relation; the host galaxy
  possibly hides just below the F606W detection limit \citep{maga05}.
  
\item The host galaxy is actually absent, \he\ is a truly ``naked'' QSO,
  by means of a black hole ejection event in a gravitational three body
  interaction or gravitational recoil following the merger of \he\ with
  the companion galaxy \citep{hoff06,haeh06,bonn07}.

\item The original black hole mass estimate was too high
  \citep{merr06,kim07,leta07} and is in fact $\sim$10 times lower. With
  comparably narrow ($\sim$1500~km/s FWHM) broad QSO emission lines the
  QSO could be the high-luminosity analog of the class of narrow-line
  Seyfert 1 galaxies (NLSy1). The host galaxy could be normal for the
  black hole mass and be absolutely consistent with the ACS upper limits.

\end{enumerate}

\noindent
In this article we present new data initially motivated by the still
undetected host galaxy and by the possibility that the host galaxy might
be obscured by substantial amounts of dust. We want to investigate the
overall cool and warm dust properties of the system, using new near
infrared (NIR) and mid infrared (MIR) images. The F606W ACS band is
strongly susceptible to dust attenuation, and dust could have prevented
the detection of the host galaxy in the optical. With new NIR data we look
at a substantially more transparent wavelength.

At the same time the new infrared data is meant to localize the source(s)
of the ULIRG emission. Three components are candidates for this: The AGN
nucleus, the host galaxy, and the companion galaxy. Our NIR data allow to
trace star formation and the MIR image traces the hot dust in the
system. We present the new data and interpret it in the view of the so far
collected knowledge from X-ray to radio-wavelengths that was built up
since the article of \citet{maga05}.
\smallskip

Throughout we will use Vega zero-points and a cosmology of
$h=H_0/(100\mathrm{km s^{-1} Mpc^{-1}})=0.7$, $\Omega_M=0.3$, and
$\Omega_\Lambda=0.7$, corresponding to a distance modulus of 40.84 for
$z=0.286$ and linear scales of 4.312 kpc/\arcsec.


\section{The IR angle}
Up to now the only existing infrared observations on \he\ were from the
2MASS survey in the near infrared $J$, $H$, and $K$ bands at $\sim$4\arcsec\
resolution, and in the MIR from the IRAS mission \citep{grij87,low88} at
12, 25, 60, and 100 $\mu$m with about 4\arcmin\ resolution. Both surveys
do not resolve the different individual components of the system (QSO,
companion galaxy, foreground star).
De Grijp et al.\ (\citeyear{grij87}) noted that the \he\ system is showing
the MIR/FIR luminosities of a ULIRG system, but it was not clear which
components of the system are responsible for this emission due to the
coarse IRAS resolution. We want to localize the dust emission in two ways:
(a) A direct observation of the hot dust component at 8.9$\mu$m
(rest-frame) with the VISIR imager at the ESO VLT. (b) A localization of
dust in general by combining new HST near infrared and the existing ACS
optical data. For this purpose we obtained HST NIC2 imaging in the
rest-frame $J$-band at $\sim$1.3$\mu$m.

\subsection{VISIR 11.3$\mu$m imaging data}
\begin{figure}
\centerline{\includegraphics[width=\columnwidth, bb=79 370 535 697]{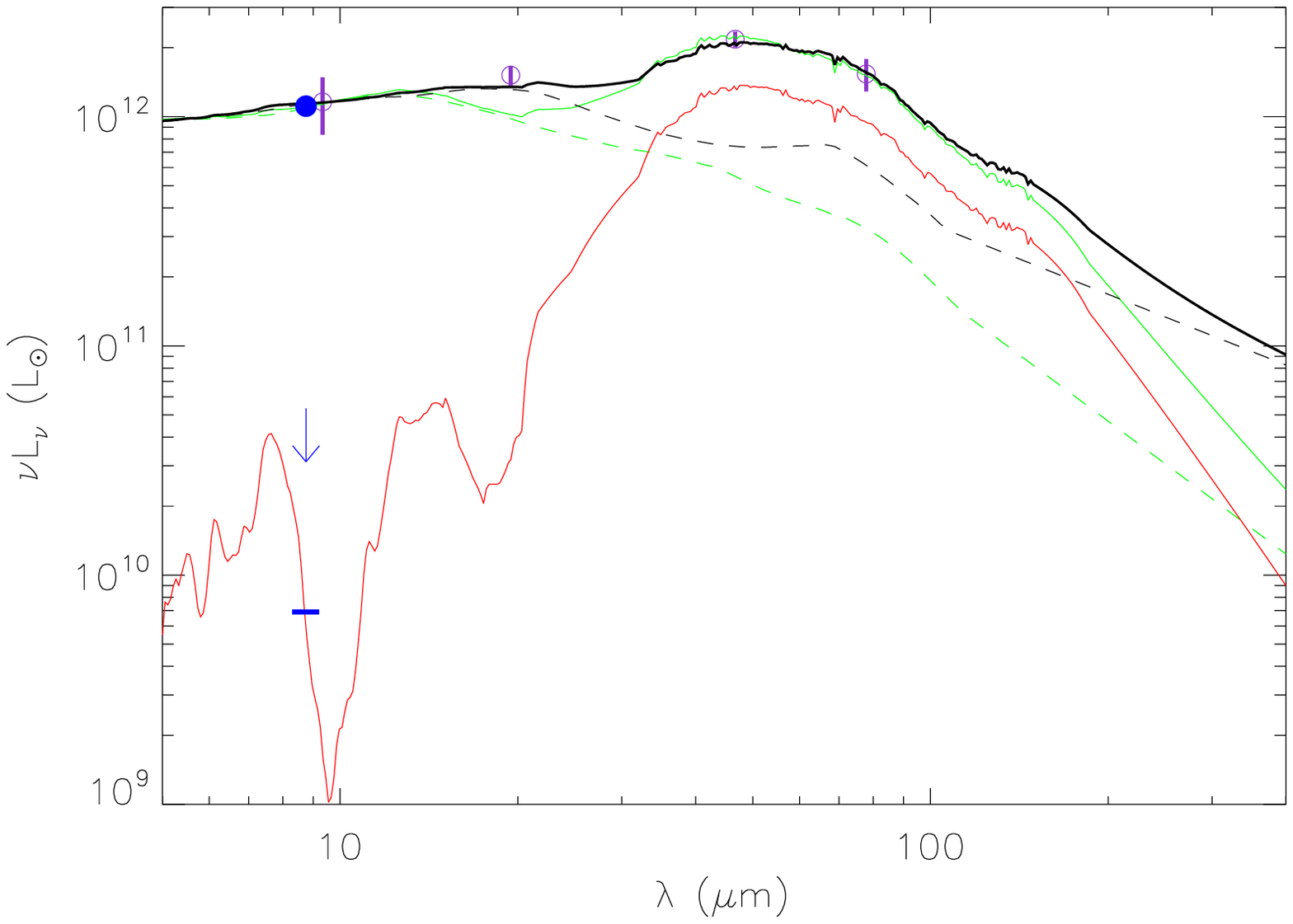}}
\caption{\label{fig:iras_sed} 
The SED of \he\ in the mid-infrared: Shown are the IRAS flux density
measurements from \citet{grij87} {\em (open circles)}, our VISIR data point
{\em (filled circle)} and upper limit on the companion galaxy {\em (arrow)}
and overlaid composite AGN plus starburst SEDs {\em (lines)}. For the quasar
nucleus we use the median {\em (green dashed line)} and 68 percentile reddest
SEDs {\em (black dashed line)} from \citet{elvi94}, the starburst {\em (red
solid line)} is a model for Arp220 by \citet{elba02}. The median quasar plus
Arp220 SED {\em (green solid line)} can explain the data except for a slightly
too low value at the observed 25$\mu$m point, but with the 68 percentile SED
{\em (black solid line)} the match is perfect. The predicted flux of the
companion galaxy where the star formation of the system is located {\em (bar)}
lies below our detection limit, consistent with the data. Milder star
formation templates as e.g.\ M82 can be ruled out, since they predict too high
fluxes for the companion -- also from the observed $H$-band data -- which
should be visible in the VISIR image.
}
\end{figure}

In the near and mid infrared the \he\ system clearly has a spectral energy
distribution (SED) that is composed of more than a single component: In
Figure~\ref{fig:iras_sed} we model the IRAS and VISIR flux densities with
a composite SED of a quasar plus a star forming component. For the quasar
we test the median and 68 percentile reddest quasar SED from
\citet{elvi94}, and for the star forming component an Arp220-like
starbursting SED, but we also tried a medium star formation M82 SED, both
from \citet{elba02}. The median quasar SED plus Arp220 can reproduce the
data at all wavlengths, except at observed 25$\mu$m, where it leaves a
small mismatch. The 68 percentile reddest SED on the other hand creates a
perfect match also there. For both cases the flux predicted for the
companion galaxy at 11.3$\mu$m lies below the detection limit as
observed. Milder, M82-like star formation can be ruled out on the same
basis, as it predicts a detection of the companion also at 11.3$\mu$m --
both with the information from the mid infrared, as well as when
extrapolating the observed $H$-band flux.

\citet{papa08} match a simple model of two black-body emission curves to
the four IRAS points, yielding a cool dust component heated by star
formation and a warm dust component which can be attributed to intense AGN
emission (see Section~\ref{sec:discussion_ulirg}). While it is not
possible to spatially resolve the system at FIR wavelengths with current
telescopes, we aim for the highest wavelength where this is currently
possible, in order to localize the warm emission component and test
whether this comes solely from the (optically visible) QSO or from extra
sources.

The observations were performed using VISIR, the ESO/VLT mid-infrared
imager and spectrograph mounted on unit 3 of the VLT (Melipal). VISIR
gives a pixel size of 0\farcs075 and a total field-of-view of
19\farcs2. The diffraction limited resolution is 0\farcs35 FWHM. Standard
``chopping and nodding'' mid-infrared observational technique was used to
supress the background dominating at these wavelength. All the
observations were interlaced with standard star observations of HD~29085
(4.45~Jy) and HD~41047 (7.21~Jy). The estimated sensitivity was 4
mJy/10$\sigma$/1h.

Imaging data were obtained on the 12th of December 2005 in service
observing mode, through the PAH2 filter centered on 11.3 $\mu$m having a
half-band width of 0.6 $\mu$m. Weather conditions were very good, optical
seeing was below 1\arcsec, and the object was observed always at an
airmass of 1.15, which resulted in a diffraction limited image of
0\farcs35 resolution.  Chopping/nodding parameters were 8\arcsec/8\arcsec\
and 0.25~Hz/0.033~Hz.  The total time spent on-source was 1623~s.  The
data were reduced using a dedicated pipeline written in IDL, which does
the chopping/nodding correction and removes the spurious stripes due to
detector instabilities \citep{pant07}.  The reduced data were finally
flux-calibrated using the two reference stars as photometric
calibrators. The error on the photometry due to variations of the
atmospheric transmission are estimated to be less than 2\% (3$\sigma$).


\subsection{NICMOS $H$-band imaging data}
The ACS $V$-band is too blue to penetrate any substantial amount of
dust. With the scenario of a dust enshrouded host galaxy in mind, we
acquired new HST NICMOS data (NIC2 with 0\farcs075 plate scale) in the
F160W $H$-band (program \#10797, cycle 15) to reduce the dust attenuation
by a factor of 3.5 in magnitude space.

A total of 5204s integration on target was forcedly split into two
observation attempts due to telescope problems, and carried out in July
2006 and 2007.  These yielded two sets of data with 2602~s integration
each, but slightly different orientations. In order to minimize chromatic
effects, we also observed a point spread function (PSF) calibrator star
(EIS~J033259.33--274638.5) with the SED-characteristics over the F160W
filter bandpass similar to a mean QSO template. We do not know the actual
SED of \he\ itself, as no NIR imaging or spectroscopic data of the system
with high enough spatial resolution exist to date. As the stellar type
yielding the likely most similar PSF we found K4III, by comparing the PSFs
predicted by the TinyTim package \citep{kris03}. The only cataloged stars
faint enough to not immediately saturate were observed by the ESO Imaging
Survey \citep[EIS,][]{groe02} located in the E-CDFS, and had to be
observed at 6 months distance in time to \he. Since we also want to
minimize the PSF variation due to differences in observing strategy, we
applied the same dither patterns for both \he\ and the PSF star. Due to
the absolute pointing accuracy of HST the centroid location of the star
relative to the chip is shifted about 15 pixels (1\farcs1) from the QSO
centroid towards the companion galaxy.

Data reduction and combination of the individual frames were carried out using
a mix of STScI pipeline data products, pyraf, and our own procedures in
MIDAS and Fortran. The resulting image is shown together with the analysis in
Figure~\ref{fig:dataimages}a. Two parts of the team analyzed the combined
images in complementary ways, by decomposition of the components using
two-dimensional modeling and by image deconvolution.

\subsubsection{Uncertainty in the PSF}\label{sec:psfuncertainty}
In order to detect a putative faint host galaxy underneath the bright QSO
nucleus we require a precise knowledge of the PSF. The PSF will vary
spatially, with the energy distribution in the filter as well as
temporally, with a changing effective focus of the telescope due to
changing thermal history.

We opt for a double approach: First, we observe the separate PSF star with
the properties described in the last section (and see
Fig.~\ref{fig:dataimages}b). Secondly, we also have the foreground star
available that is located at 1\farcs8 distance from the QSO to the north-west.
It is classified as a G star \citep{low89}. Its on-chip distance to the QSO
will leave only room for small spatial variations, but its SED in the $H$-band
likely will not perfectly match the SED of the QSO.

It is difficult to assess the PSF uncertainty at the position of the
QSO. In principle we have a combined effect of color, spatial, and
temporal variation, but only one bit of information: the difference
between the foreground star and the PSF star. We thus model the expected
difference in the shape of these two stars with TinyTim and then compare
their actual observed shapes. This shows that the foreground star should
be slightly narrower than the observed PSF star, which is consistent with
PSF star's later, redder spectral type and an increase of PSF width with
wavelength. We observe this effect also in the data, however somewhat
stronger. A temporal variation can thus not be separated and ruled out.

In any case we conclude that the PSF star is wider and thus will yield
more conservative (=fainter) estimates for a QSO host galaxy, while in
case of a non-detection the foreground star will yield brighter upper
limits.

For two-dimensional modeling of the system we use {\sc galfit} \citep{peng02}.
In order to quantify the PSF uncertainty for this process, we first let {\sc
  galfit} fit a single point source, represented by the PSF star, to the
foreground star. In this process we use an error map created from the data
itself and we add the sky as a free parameter. We minimize the influence of
the nearby QSO on the foreground star by first fitting the former with a
single point source as well, removing its modeled contribution, and mask out
the remaining residuals starting at 0\farcs9 from the star. The PSF created in
this way is shown in Figure~\ref{fig:dataimages}c. This image is fed into the
modeling process of the PSF star, or later the QSO/host/companion system.

The residual flux in this process is of the order of 3\% of the total,
inside the 0\farcs5 radius aperture where most apparent residuals are
located, the absolute value of the residuals in the same region is
14\%. This means that it will be generally impossible to detect any host
galaxy of less than 3\% of the total flux of the QSO, and it will even be
difficult to isolate a somewhat brighter smooth galaxy in the non-smooth
residuals. This level of residuals is consistent with experience from the
HST ACS camera, where we find that due to PSF uncertainties 5\% of the
total flux are approximate detection limits for faint host galaxies
\citep[Jahnke et al.\ in prep.]{jahn04b}. Including the structured PSF
residuals we will only consider a host galaxy component as significant
if it has clearly more than 3--5\% of residual flux inside an 0\farcs5
radius of the QSO, or that shows up as a non co-centric structure above
the noise outside this region.

In absolute magnitudes and related to the QSO these limits correspond to
the following: inside an 0\farcs5 radius of the QSO we can hide a galaxy
co-centric with the QSO of at least $M_H\sim-24.7$ (for the 3\% case) or
$M_H\sim-25.2$ (for 5\%).

\section{Results}
\subsection{VISIR}

We detect a single unresolved point source in the VISIR field-of-view with
a flux density 62.5~mJy at observed 11.3$\mu$m (Figure~\ref{fig:allwave},
right column). This compares to 69.3~mJy in the IRAS 12~$\mu$m
channel. There is no second source detected in the field down to a point
source sensitivity of at least 3~mJy at the 5$\sigma$ level. Extended
sources of the visual size of the companion galaxy have a 5$\sigma$
detection limit of 5.5~mJy.

With only one source in the total 19\farcs2 VISIR field three optical
sources have in principle to be considered as potential counterparts: The
QSO nucleus, the companion galaxy, and the foreground star. However, the
star is a G spectral type and can thus be safely ruled out.

We find that the initial position of the MIR point source as recorded in
the VISIR image header comes to lie between the QSO and the companion,
somewhat closer to the QSO. To clarify this we conducted an analysis of
the pointing accuracy of VISIR testing the astrometry of a number of
reference stars observed with VISIR at different epochs. The two results
are: (1) In all cases the offset between targeted and effective RA,Dec is
less than 1\arcsec\ rms, but (2) there is a systematic offset of 0.15s in
RA recorded in the fits header, so the true positions need to be corrected
by --0.15s in RA. This correction places the MIR point source exactly onto
the locus of the QSO in the HST ACS images. It is thus clearly the QSO
nucleus that is responsible for all of the 11.3~$\mu$m emission.

\subsection{NICMOS}
\subsubsection{Host galaxy}\label{results:host}
To extract information on the host galaxy, we use three different methods
to remove the flux contribution from the QSO nucleus. First, we make a
model-independent test for obvious extended emission: In a simple peak
subtraction we remove a PSF from the QSO, scaled to the total flux inside
two pixels radius around the QSO center. This is a robust approach that is
independent of specific model assumptions and quite insensitive to the
noise distribution in the image \citep{jahn04b}. As a result, the peak
subtracted image shows no obvious extended residual, i.e.\ host galaxy,
centered on the QSO, when using the PSF star as PSF.

As a second step we use on the one hand {\sc galfit} to model the
2-dimensional light distribution of the \he\ system and decompose it into
different morphological components. On the other hand we use the MCS
deconvolution method \citep{maga98} to mathematically deconvolve the
system to a well defined and narrower PSF. The procedure we follow is
based on the one described in \citet{chan07}. For {\sc galfit} we use the
two empirical PSFs, for MCS deconvolution we construct a number of
combinations of empirical PSF and TinyTim models including very red
dust-like SED components.

\begin{figure*}
\begin{center}
\includegraphics[width=\textwidth,clip]{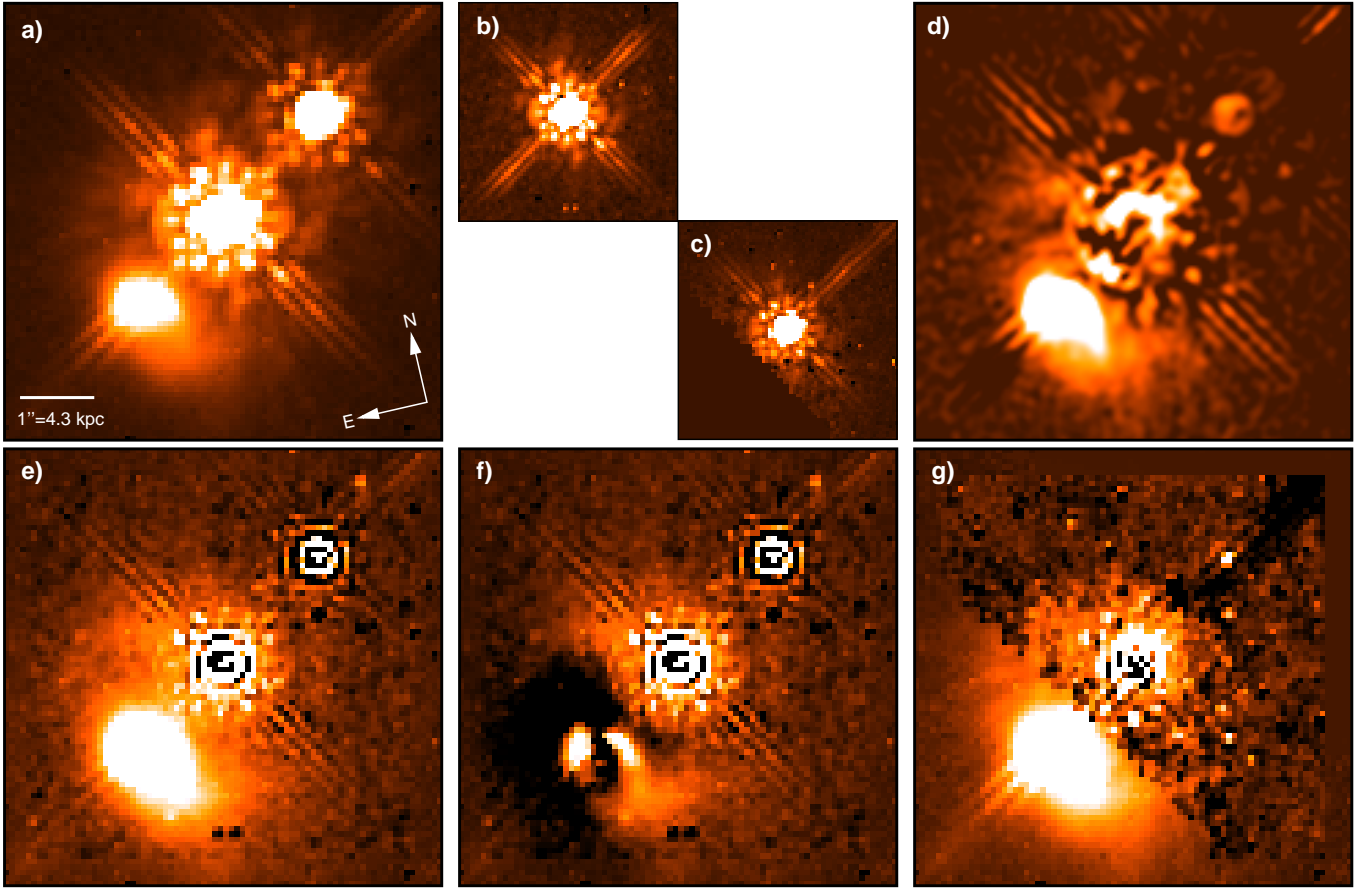}
\end{center}
\caption{\label{fig:dataimages} 
NICMOS $H$-band data. Shown are {\em (a)} original observed image, {\em (b)}
the separately observed PSF star EIS~J033259.33--274638.5 and {\em (c)}
cleaned and masked foreground star used as alternative PSF. The system is
further shown after nucleus removal: {\em (d)} System after MCS-deconvolution
and {\em (e--g)} after 2dim-modeling with {\sc galfit}. Deconvolution uses as
a PSF an envelope from the PSF star {\em (b)} with a core from the companion
star {\em (c)}. For {\sc galfit} modeling, QSO and foreground star are
represented with one point source each, using either {\em (e+f)} the PSF star
as a PSF or the {\em (g)} foreground star. The companion is neither
represented well with one nor two S\'ersic components as seen in {\em (f)} for
the otherwise same setup as {\em (e)}, where the companion was not medelled.
No significant host galaxy is seen cocentric with the QSO center. However
a faint emission is visible at $\sim$0\farcs6--1\farcs5 to the north-east
(see Figure~\ref{fig:ne-extension} for more details).
The displayed image size is 80 NIC2 pixels of 0\farcs075 size, i.e.\
6\arcsec.
}
\end{figure*}

While these two approaches are complementary in method, their results
agree as can be seen in Figure~\ref{fig:dataimages}: The inner part of the
QSO inside of 0\farcs5 radius is consistent with a point source within the
PSF uncertainties, but there is extra flux present outside of this
radius. The structure of the PSF removal or deconvolution residuals points
to a substantial mismatch between shape of the QSO nucleus and the
separately observed PSF star, but also to too simple models of TinyTim. In
order to remove obvious residual PSF structure a very red SED needs to be
assumed, which at this point can not be discriminated from a marginally
resolved red component on top of the AGN point source. However, in light
of the non-average properties of this QSO, a mean QSO SED is also not
expected.  \medskip

In the following we present our results in more detail and focus on the
{\sc galfit} results, since it allows a more direct estimate of the
significance of detected structures. A comparison of the original and point
source-removed images in the optical and NIR, and the MIR image are shown in
Figure~\ref{fig:allwave}.

We use {\sc galfit} to perform a number of different model fits. In all of
them the companion star and QSO nucleus are described by a pure point
source, while the companion galaxy is fit with one or two
S\'ersic\footnote{The S\'ersic profile \citep{sers68} is a generalized
galaxy profile with variable wing strength, set by the
``S\'ersic-parameter'' $n$.  It reverts to an exponential disk profile
typical for spiral galaxies for $n=1$ and for $n=4$ it becomes a
de~Vaucouleurs profile found for many elliptical galaxies.} components
with free axis ratio, or left unmodelled. We also attempt to add another
S\'ersic component for the putative host galaxy.  We always leave the
S\'ersic parameter $n$ free, although the companion galaxy is too complex
and the putative host galaxy too faint for $n$ to be interpreted
physically.

With the PSF star used as PSF {\sc galfit} finds a result consistent with
the peak subtraction. A positive residual of $H\sim17.7$ inside a
$\sim$1\arcsec\ radius aperture has a flux below 2\% of the 13.05~mag of
the QSO itself (see Figure~\ref{fig:dataimages}e+f). Even though we choose
an aperture larger than in our calculation in
Section~\ref{sec:psfuncertainty}, we receive a value far below our
significance limit, so no significant co-centered host galaxy is seen in
this way.

If we use the foreground star as PSF (Fig.~\ref{fig:dataimages}g) we find
-- as expected -- a residual flux that is slightly higher than before, and
consistent values for two different approaches: First, for a pure PSF fit
to the QSO location, integrating the flux of the residual within a
1\arcsec\ radius aperture around the QSO, except along the SE--NW-axis
where we expect residual flux from foreground star and companion
galaxy. Secondly, we get a similar flux for a fitted additional host
galaxy S\'ersic component.

These two approaches yield a magnitude of $H$$\sim$15.8 and 16.2,
respectively, for the host, $\sim$1.5mag brighter than for the PSF star
fit. $H$$\sim$16 corresponds to $\sim$6\% of the 13.05~mag of the QSO
nucleus.

Again the QSO residual shows substantial structure as reported in
Section~\ref{sec:psfuncertainty}. It consists of nested rings of positive
and negative flux, typical signs of a close but different width between
the PSF we use and the actual one. The bulk of structure is contained in
the innermost 0\farcs5 radius and contains 2/3 of the residual flux. The
remaining residual of 2\% of the total flux outside this radius is again
insignificant, and no main body of the host galaxy co-centered with the
quasar is found which satisfies our significance criterion.  Going back to
the PSF residuals that we quantified earlier on, we detect no co-centered
host galaxy at a level above 3\% of the flux of the quasar nucleus,
corresponding to an upper limit of $H=16.9$.

\begin{figure}
\centerline{\includegraphics[width=\columnwidth]{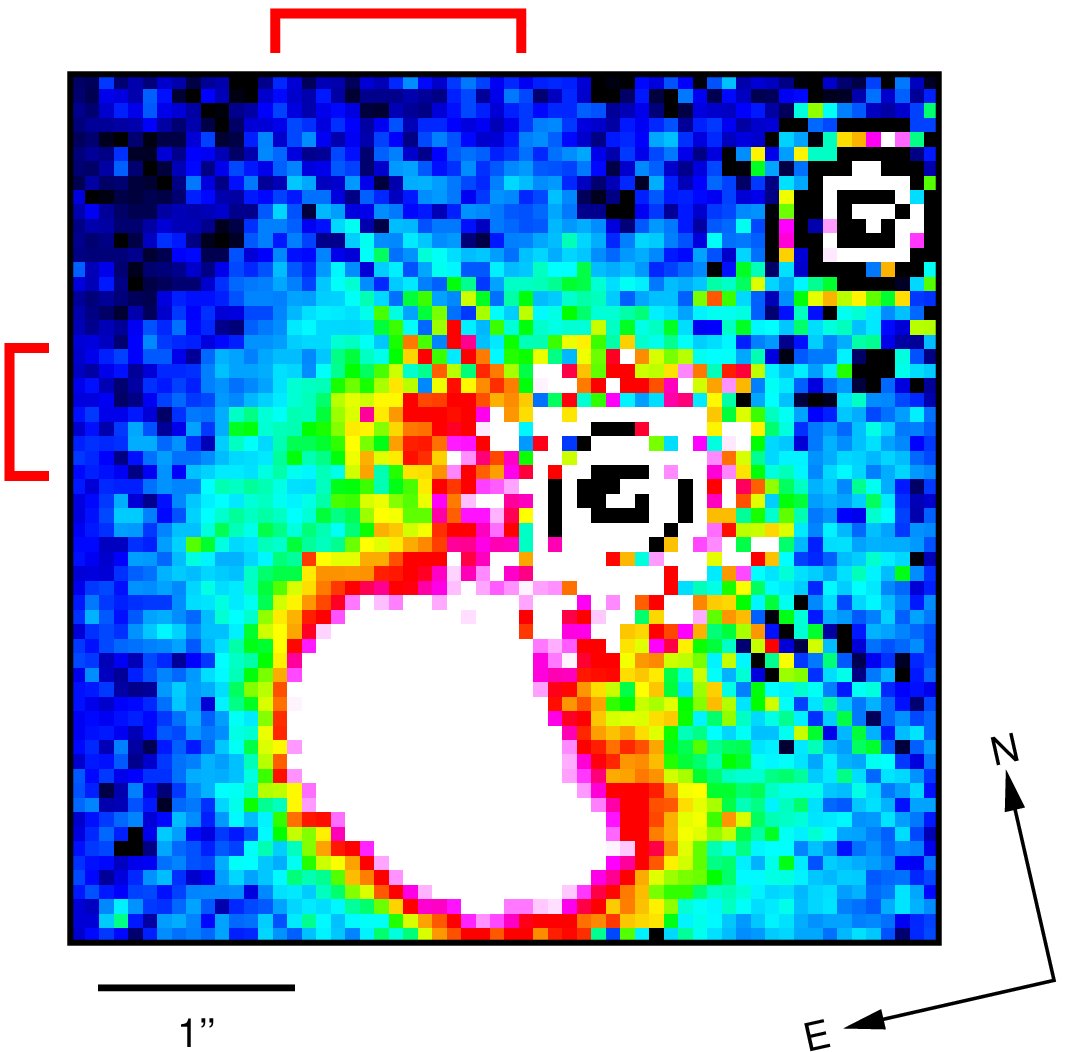}}
\caption{\label{fig:ne-extension} 
A slight zoom into the inner region of \he\ to show the newly found
``NE-extension'' of the QSO. We removed the star and the QSO using the PSF
star as PSF. An extension to the N--E is visible (marked with {\em red
brackets}) at a distance of 0\farcs6--1\farcs5 that is clearly not due to PSF
residuals -- a very similar result is seen when using the narrower foreground
star as PSF (Fig.~\ref{fig:dataimages}e), or MCS deconvolution
(Fig.~\ref{fig:dataimages}d). This structure is disjoint from the companion
galaxy so very likely belongs to the QSO host galaxy itself. The estimated
brightness is $H=18.8$. The image size is 4\farcs5 on the side.
}
\end{figure}

However, after removal of the point source, a feature becomes clearer,
what we dub the ``NE-extension''. This faint structure extends from the
QSO to the N--E, and it can be traced starting at the edge of the strong
PSF residuals at 0\farcs6 (2.5~kpc) N--E of the nucleus
(Figures~\ref{fig:dataimages} and \ref{fig:ne-extension}). Some signs of
it are already visible in the optical, when going back to the the F606W
image \citep[][see also Fig.~\ref{fig:allwave}]{maga05}, but it is much
more pronounced in the new $H$-band data compared to the $V$-band. The
NE-extension is possibly part of a tidal arm similar to the arm towards
the south of the companion, already described by \citet{cana01}, but our
$H$-band image shows it to be clearly disjoint from the companion
galaxy. Due to its proximity it is very likely associated with the QSO,
even though it is clearly not its main body. It is unlikely that the
NE-extension is just a gas cloud with star formation induced by the radio
jet in the system, since it lies at least 50$^\circ$ from the jet
direction \citep{feai07}. It is also unlikely a chance superposition of a
gas cloud with emission line gas, as seen by \citet{leta08}, since the
observed $H$-band does not contain any strong enough line. The
NE-extension contains non-negligible flux far above the noise of the
background and is unaffected by QSO residuals and independent of the PSF
used. We estimate its brightness at $H=18.8$ using an aperture
encompassing all visible extension outside the QSO nucleus residual. The
same region in the ACS image has $V=21.6$, so $(V-H)=2.8$

In summary, we detect no significant host galaxy that is co-centered with
the QSO. We conclude this from the size and shape of the residuals
underneath the QSO in comparison to the ``PSF star minus foreground star''
subtraction residuals we discussed above. The NE-extension, however, that
can be seen outside of the residuals of the QSO nucleus, is a real and
significant emission structure -- and it is very likely associated with
the main part of the host galaxy.

\subsubsection{Companion galaxy}
\begin{figure}
\centerline{\includegraphics[width=\columnwidth]{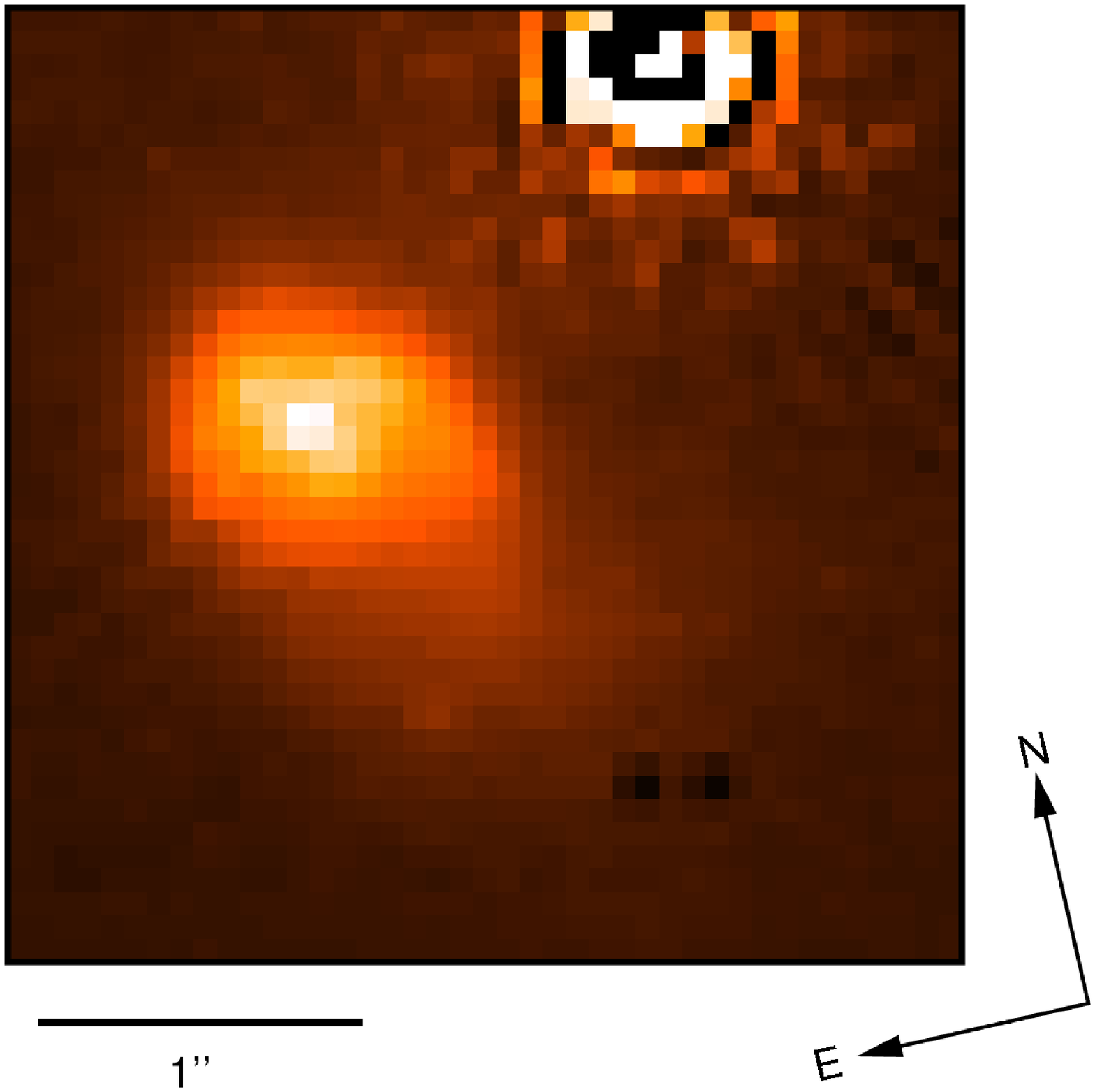}}
\caption{\label{fig:companion} 
Zoom on the companion galaxy. As a difference to the $I$-band
(Fig.~\ref{fig:allwave}) the galaxy has a pronounced peak of emission and
no ring. The light in the optical is obviously attenuated by dust, very
strongly in the center where the dust is optically thick, less in the
outer regions. Image size is 3\arcsec\ on the side.
}
\end{figure}

In the ACS $V$-band the companion galaxy located 1\farcs5 to the S--E
appears clumpy, with several bright knots as well as lower surface
brightness in the center.  \citet{cana01} even call the companion a
``collisional ring galaxy''. With the NICMOS $H$-band we get a
substantially different picture. The galaxy at $H=15.2$ is still
asymmetric, with tidal extensions, but contrary to the visible wavelengths
it is smooth and shows a pronounced center: clear signs for substantial
dust, distributed not smoothly but unevenly and clumpily, with
concentration towards the center that only shows up in the optical
(Fig.~\ref{fig:companion}). The complexity of the companion is manifested
in that there is no good description with neither one or two Sersic
components, when the azimuthal shape is restricted to ellipses. The Sersic
index of the companion is around $n=2$ for a single Sersic component, and
$n<1$ if two Sersic components are used. Taken at face value, both cases
point to a more disk- than bulge-like companion, but a substantial
fraction in flux is containted in the non-symmetric distorted part of the
companion -- and this should be the main description of the companion.
More complex descriptions were put forward, with either a proposed
additional faint AGN hosted by the companion galaxy \citep{leta09},
explosive quasar outflows \citep{lipa09}, or quasar-induced star formation
\citep{elba09}.

\section{Discussion}

\subsection{Where is the ULIRG?}\label{sec:discussion_ulirg}
There was substantial confusion about the source for the ULIRG-strength
IRAS MIR and FIR emission in the literature. From the uncorrected [OII]
line flux a star formation rate (SFR) of only 1~M$_\odot$/yr can be
inferred \citep{kim07}. \citet{maga05} still assign the ULIRG emission to
the companion galaxy due to its Balmer decrement which yields
non-negligible dust extinction, while \citet{kim07} note that the
corrected SFR would still be below 10~M$_\odot$/yr. This number is in
strong disagreement with a SFR up to $\sim$800~M$_\odot$/yr inferred
from total IR-luminosity or 370~M$_\odot$/yr from CO \citep{papa08}.

The new NICMOS images show that the stars in the companion galaxy are not
distributed in a ring, but smoothly (Fig.~\ref{fig:companion}) and that an
optically thick dust creates the ring-like structure in the optical ACS
images (Fig.~\ref{fig:allwave}a+b). This means that at optical wavelengths
only information from the less extincted outer regions of the galaxy as
well as the surface of the strongly extincted central regions is
seen. UV-based SFRs must therefore dramatically underestimate the true
SFRs when corrected with dust extinction estimated from (also optical
wavelength) Balmer-decrements.

The actual scale of the uncertainty in $A_V$, the optical extinction
correction, can be estimated by comparing $A_V$ estimates from Balmer
lines and Paschen/Bracket lines in other ULIRGs. \citet{dann05} studied
five ULIRGS for which they estimated $A_V$ both from H$_\alpha$/H$_\beta$
as well as from Pa$\alpha$/Br$\gamma$. NIR-derived values for A$_V$ were in
every case significantly larger, ranging from factors of $\sim$1.16 to
$\sim$10 (mean 4.0) times higher. As this factor does not scale in any way
with the optical $A_V$ estimate, but only with the NIR estimate, we can
not determine a correction for \he.

When starting out with the redshifted [OII]-line at $\lambda$3727 and
$A_\mathrm{3727}=A_V\times1.57$ like \citet{kim07}, and the correction
factors from \citet{dann05}, a huge range of possible star formation rates
arises.  An average of the two {\em lowest} correction values from
\citet{dann05} of 1.16 and 1.65 means $A_V\sim 2.1$,
$A_\mathrm{3727}\sim3.3$ or corrected SFRs of 21 M$_\odot$/yr. Using their
mean correction factor of $\sim$4 would lead to $A_\mathrm{3727}\sim9.4$
or $>$5000 M$_\odot$/yr. So already a number below the mean correction
($A_V\sim3$) would make these numbers consistent with FIR-emission based
SFR estimates. This directly shows that optical/UV line-emission based
SFRs as used by \citet{kim07} can not at all be used to constrain the true
SFR of ULIRGs and does not provide an argument against strong star
formation in the companion.
\smallskip

\citet{papa08} approximate the IRAS IR SED with a 2-component black-body
model and find a cool component $T_\mathrm{dust}^\mathrm{cool}=47$K, dust
mass $M_\mathrm{dust}^\mathrm{cool}\sim10^8$ M$_\odot$ and $L_\mathrm{FIR}
\sim 2.1\times10^{12}$ L$_\odot$, and a warm component with
$T_\mathrm{dust}^\mathrm{warm}=184$K,
$M_\mathrm{dust}^\mathrm{warm}\sim5\times10^4$ M$_\odot$ and
$L_\mathrm{MIR} \sim 2.6\times10^{12}$ L$_\odot$. We can now for the first
time spatially localize the warm component from the detection of the
single 11.3$\mu$m point source with VISIR to be coincident with the
position of the QSO nucleus. Since the measured flux density is consistent
with a warm component having the previously known 12$\mu$m IRAS flux
density, we conclude that the QSO nucleus itself already is a ULIRG-level
emitter, but with a warmer component compared to star formation.

For localizing star formation in the system, there are two recent new
datasets available, radio data from \citet{feai07} and the CO maps by
\citet{papa08}. While the radio maps do not set strong constraints when
trying to exploit the radio--FIR relation to assign a location for the FIR
emission, the CO data are more powerful: at least the bulk, possibly all
of molecular gas and thus star formation activity is located in the
companion galaxy.

We can add two further constraints from our NICMOS and VISIR images. Both
the mid infrared SED of the system (Fig.~\ref{fig:iras_sed}) as well as an
extrapolation from the $H$-band are consistent with an Arp220-like star
formation, while ruling out milder, M82-like conditions. In the latter
case the companion would have to be visible in our observed 11.3$\mu$m
image, but it is absent (Fig.~\ref{fig:allwave}). Together with the dense
and clumpy dust geometry of the companion when comparing optical and NIR
morphology, it becomes clear that the companion is responsible for most,
if not all, of the 370~M$_\odot$/yr star formation.

If we follow the 5:1 CO detection significance for the companion given by
\citet{papa08}, this means that as a minimum 5/6=83\% of CO are located in
the companion and thus also $\ge83$\% of the star formation and FIR
emission. This number converts to an integrated IR luminosity of
$L_\mathrm{FIR} \ge 1.75\times10^{12}$ L$_\odot$, so the companion also
qualifies as a ULIRG.
\smallskip

While the presence of very strong star formation in the companion is clear
now, its trigger is a priori not so clear. The most probably solution is
merger induced SF, so the system would be a classical ULIRG -- just with a
non-standard geometry -- but there is room for a radio jet induced effect
as well. One of the lobes of the jets from the QSO is located directly at
the companion position. If and how much this contributes to star formation
in the companion still needs to be quantified.

\subsection{Host galaxy detection}\label{sec:hostgalaxydetection}
With the companion identified as the main star-former, we get limits from
the CO that less than 1/6 of the total cool dust is located within the
putative host galaxy. Thus 1/6 of the FIR-inferred SFR by \citet{papa08}
of $\mathrm{SFR} = 1.76\cdot 10^{10} (\mathrm{L_{IR}/L_\odot})$
M$_\odot$/yr correspond to an upper limit of 62~M$_\odot$/yr. This leaves
room for a non-negligible amount of SF in the host galaxy, but is also an
upper limit\footnote{Note that for the galaxy-scale star formation regions
around QSO nuclei the dust can be heated by a mix of stellar emission as
well as energy from the AGN. In this sense the 47~K found for the cool
dust component of \he\ agrees well with the mean SF-heated dust around
higher-$z$ QSOs \citep[also 47~K,][]{beel06}, and can be composed of
intrisically cooler dust (20--30~K) plus AGN heating. This temperature
could thus be a hint that indeed a part of this cool dust component is
located in the QSO host galaxy and not in the companion.}.
If we assume the host galaxy to have a mix of old and young stellar
population as we find for other QSO host galaxies at these redshifts
\citep{jahn04a,leta07}, we can convert this to an expected $H$-band
flux. If the host galaxy had the same population mix as \citet{cana01}
modelled for the companion galaxy\footnote{\citet{cana01} used optical
spectra only. With the optically thick dust now detected we have to
restrict their diagnosis to mainly the outer parts and surface of the
companion. The population mix there might be identical to the core of the
companion, but it does not necessarily have to.}  -- 95.5\% of a 10~Gyr
old population with 5~Gyr e-folding SFR timescale plus 4.5\% of a 128~Myr
young population --, this SFR upper limit would translate to an expected
NIR magnitude of 1.75~mag fainter than the companion or $H\ge16.95$. The
combined color and $K$-correction term is $V-H_\mathrm{z=0.285} = 1.66$,
and changes by only about $\pm$0.3mag for a pure old (10~Gyr) or young
(100~Myr) population. So they are rather insensitive to the exact choice
of stellar population. However, this limit will get brighter if the host
galaxy contained less dust -- by about 0.3~mag per magnitude decrease in
$A_V$.

With that in mind, this limit is not more stringent than the limit from
NICMOS itself: No significant main host galaxy body is found after PSF
removal (Section~\ref{results:host}) and so an upper limit from the NIR
decomposition of $H=16.9$ applies for a host galaxy co-centered with the
quasar nucleus. We therefore conclude that the current upper limit from
NICMOS lies at around $H\sim16.9$. This is consistent with the CO/FIR
limits.

How do these numbers relate to the current upper limit for a co-centered
host galaxy from the optical HST data?  We convert our $H$-band limit to
absolute $V$-band magnitudes with again the assumption of the host galaxy
having the same stellar population mix as the companion. In the conversion
to $M_V$ we assume two different values for dust extinction, (a) $A_V=0$,
motivated by the nearly dust-free line of sight to the QSO nucleus, and
(b) a moderate $A_V=1$ (corresponding to
$A_{H\mathrm{(z=0.285)}}\sim0.29$). This yields host-galaxy upper limits
of $M_V>-21.25$ and $>-22.55$, for the cases (a) and (b) respectively. If we
convert the \citet{maga05} upper limits to our $h=0.7$ cosmology and
assume the same stellar population and dust properties, we receive
$M_V>-20.6$ and $>-21.6$, respectively. We note here that this corresponds
to a detection limit of only 1.5\% of the total quasar flux in the
optical. This factor of two is owed to the better determined PSF in the
ACS images. This allows \citeauthor{maga05} to set somewhat stricter upper
limits for a nucleus co-centered host galaxy component, particularly if a
low dust extinction is present.  \smallskip

Concerning lower limits to the host galaxy, the NE-extension
(Figure~\ref{fig:ne-extension}) is a structure of real emission that can
be traced towards the QSO from $\sim$1\farcs5 to a radius of 0\farcs6,
where the region of substantial PSF residuals begins. We can not say for
sure whether it continues further inward from this position. Signs of this
structure are visible in the ACS $V$-band (see Figure~\ref{fig:allwave},
left column) but it is not clear whether the more compact region only
$\sim$0\farcs2 N--E of the nucleus in ACS image is real or an artefact of
the deconvolution process. We measured the $(V-H)$-color to be 2.8 outside
this region, which is consistent with a stellar population of
intermediate age. In the dust-free case this color corresponds to a
$\sim$2.1 Gyr old single stellar population \citep[][solar
metallicity]{bruz03}, for $A_V=1.0$ to an age of 800 Myr. This is
consistent with stellar material from a host galaxy, e.g.\ tidally ejected
disk stars.

We conclude that with its spatial detachment from the companion galaxy
this NE-extension is likely a part of the host galaxy, possibly as a tidal
extension, but its vicinity to the QSO makes other interpretations less
likely. With this interpretation, we receive an $H\le18.8$ {\em lower}
limit for the host, corresponding to $M_V<-20.4$ ($A_V=0$) or $<-20.7$
($A_V=1$). If we include this off-center emission to the upper limit of a
co-centered host galaxy, we obtain a total host galaxy upper limit of
$M_V>-21.2$ and $-22.0$. We thus bracket the host galaxy luminosity in the
$V$-band by 0.8 and 1.3~mag or factors of $\sim2$ and $\sim3.5$,
respectively.
\smallskip

Formally, the CO detection significance and NICMOS give the same limit on
a star formation rate of up to $\sim60$~M$_\odot$/yr. If we take into
account the stricter ACS $V$-band limits of $M_V>-20.6$ and $>-21.6$,
depending on dust cases (a) and (b), these are fainter by 1.3 and 0.6mag
than the CO predicted magnitures. Inversely, these reduce the upper limits
on star formation to 18 and 35~M$_\odot$/yr, respectively. Beyond
$A_V=2$mag the CO and NICMOS limits again become the most stringent. This
means that we can not rule out dust obscuration in the host galaxy. At the
same time the dust-free line of sight to the quasar nucleus is a strong
argument against large amounts of dust, unless a very special geometrical
configuration is invoked, while the warm ULIRG emission from the QSO
points to dust in the very central few 100~pc. Only better CO limits or a
detection of the host galaxy in the NIR will be able to finally resolve
this matter.

\subsection{Black hole mass, galaxy luminosity, and the NLSy1 angle}
Black hole mass estimates for \he\ vary significantly through the
literature. The original 8$\times$10$^8$~$M_\odot$ \citep{maga05} were
revised later to a substantially lower value of 4$\times$10$^7$
\citep{leta07}. Both values are virial estimates based on H$\beta$ width,
but while narrow and broad components were separately measured in the
former study, the FWHM of the whole line was used in the latter. This
revised value is consistent with the independent virial estimate of
6--9$\times$10$^7$ by \citet{merr06}, and even with an estimate from X-ray
variability, $2^{+7}_{-1.3}$$\times$10$^7$ \citep{zhou07}. Since the
virial estimates agree now, we will adopt the range
4--9$\times$10$^7$~$M_\odot$ for the black hole mass.

\citet{merr06} noted the rather narrow broad emission lines of \he\ and
suggested that it should actually be viewed not as a standard QSO but as a
higher-$L$ analog of local NLSy1s. If we compare \he\ with estimates from
the literature \citep{grup04,ohta07}, we find that \he\ is consistent with
the high black hole mass end of the known NLSy1 distribution and does not
need to constitute a new ``higher-$L$ NLSy1 analog'' class of its own. But
is it consistent regarding other properties as well?

Morphologically, NLSy1 are mostly spirals, often barred, mostly not
strongly disturbed \citep{ohta07}. Since galaxies have increasing bulge
mass with increasing black hole mass it is not clear which structural
properties to expect and if a merging system like this is consistent with
the properties of the local, lower mass NLSy1 population.

There is even a debate on how different NLSy1 actually are from normal
Seyferts.  Recent studies show smaller BH mass differences between normal
broad-line Sy1 and NLSy1 when using line dispersions instead of FWHM
\citep{wats07}, although a difference might remain. If galaxies with
potentially core outflow-affected lines are considered separately, NLSy1
share the same $M_\mathrm{BH}-\sigma_\mathrm{bulge}$-relation with BLSy1,
but their accretion rates are confirmed as lying often close to the
Eddington limit \citep{komo07}. If we compute the \he\ accretion rate --
as derived from the $V$-band absolute magnitude of the quasar nucleus
($M_V=-25.75$, recomputed from the HST/ACS data with updated AGN color and
$K$-correction) and a bolometric correction of $BC_V\sim8$
\citep{marc04,elvi94} -- in relation to its Eddington accretion rate, we
obtain from $M_\mathrm{BH}=6.5\pm2.5\times10^7$~$M_\odot$ a super-Eddington
accretion rate of $L/L_\mathrm{Edd}=6.2^{+3.8}_{-1.8}$. This is consistent
with high Eddington ratios observed for NLSy1 \citep{warn04,math05a}.

\begin{figure}
\centerline{\includegraphics[width=\columnwidth,clip]{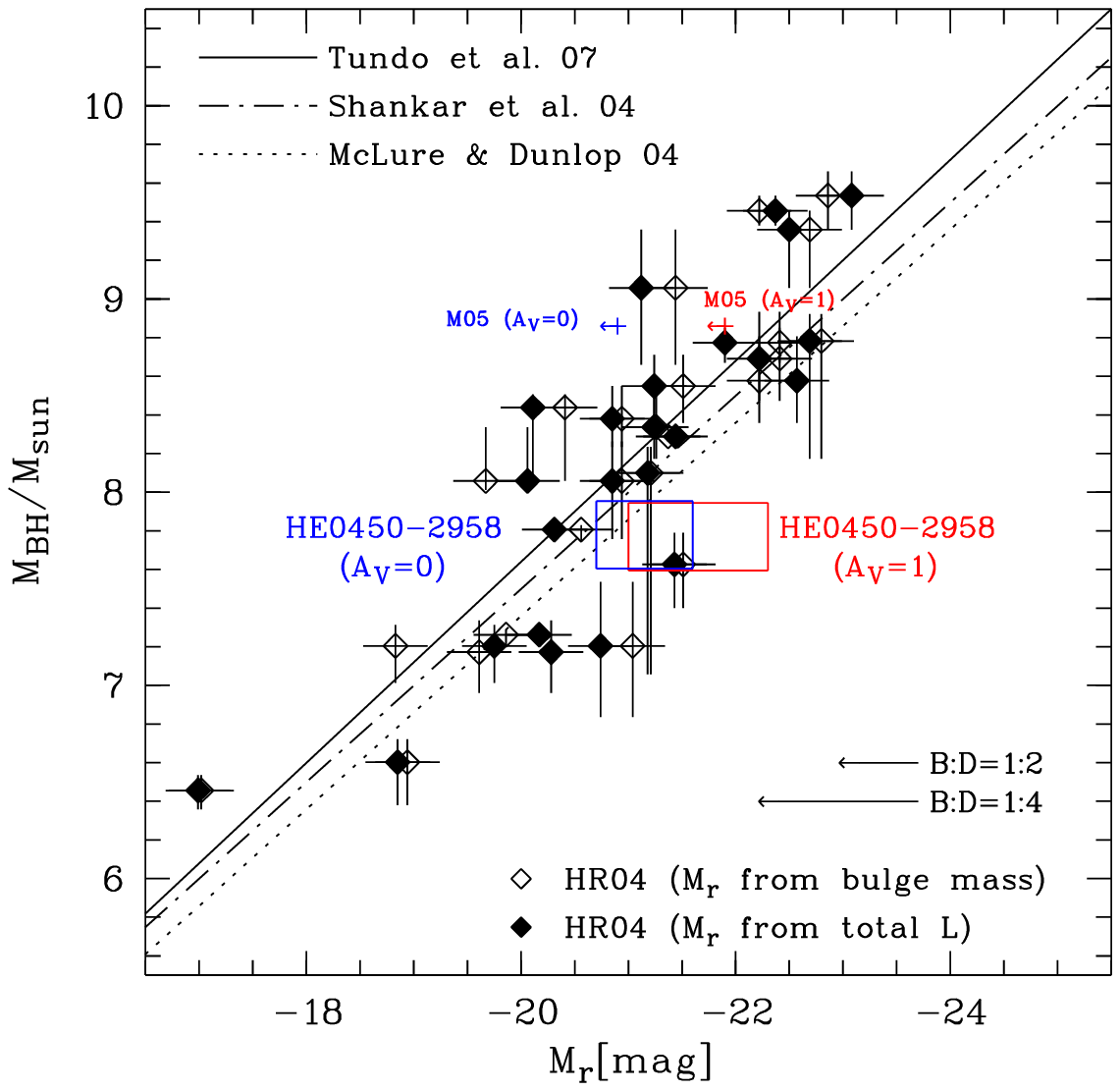}}
\caption{\label{fig:m_m} 
$M_\mathrm{BH}$--$L_\mathrm{bulge}$-relation for inactive galaxies in the
  local Universe as presented by \citet{tund07}, with data from
  \citet{haer04}, \citet{shan04} and \citet{mclu04} {\em (black lozenges and
  lines)}. Overplotted are the upper limits for the host galaxy of \he\ for
  the dust-free case by \citet{maga05} from $V$-band imaging {\em (small blue
  arrow)} and with an $A_V=1$ added {\em (small red arrow)}, with their
  original black hole estimate, converted to our cosmology. The {\em blue and
  red rectangles} show the range for black hole mass estimates and our new
  lower limits for the (total) galaxy luminosity from NICMOS and new upper
  limits based on the (still better constrained) optical HST data. Note: Here
  we combined the off-center flux lower limit (NICMOS $H$-band) with the upper
  limit for a co-centered host galaxy (ACS $V$-band) for a total upper
  limit. The arrows to the bottom right show the conversion of our
  $L_\mathrm{galaxy}$ limits to $L_\mathrm{bulge}$ limits for bulge-to-disk
  ratios of 1:2 and 1:4. Both the dust-free as well as the $A_V=1$ dust case
  show a galaxy that is absolutely consistent with the black hole mass, even
  if the bulge-to-disk ratio is accounted for.
}
\end{figure}

With the new data and an explicit assumption/interpretation that the
NE-extension is indeed associated with the host galaxy, we can for the
first time present a black hole mass for \he\ and bracketing limits for
its host galaxy luminosity. We can thus place \he\ on the
$M_\mathrm{BH}$--$L_\mathrm{bulge}$-relation of active and inactive
galaxies, with more than just an upper limit for galaxy luminosity.  In
Figure~\ref{fig:m_m} we show data from \citet{haer04} and others, as
collected by \citet{tund07}. We overplotted the limits on \he\ for the two
assumptions of dust attenuation strength
(Sec.~\ref{sec:hostgalaxydetection}).  This shows that even when applying
a sensible conversion factor of 1 to 1/4 (up to 1.5~mag) to convert from
total to bulge luminosity, the host of \he\ will be a perfectly normal
galaxy in this parameter space, with a luminosity around the knee of the
galaxy luminosity function, $L\sim L^*$.

Contrary to the claim by \citet{maga05} it does not deviate substantially
from the local $M_\mathrm{BH}$--$L_\mathrm{bulge}$-relation for normal
inactive local massive galaxies, mainly due to the revised mass estimate
for the black hole.  However, this also means that \he\ does not show a
$M_\mathrm{BH}$/$L_\mathrm{bulge}$ different from local broad-line AGN,
consistent with being a NLSy1-analog if the \citet{komo07} result is taken
as a base.

With the normal $M_\mathrm{BH}$/$L_\mathrm{bulge}$-ratio and the fact that
we can now rule out huge amounts of obscuring dust around the QSO nucleus,
the most likely explanation for the evasive host galaxy is indeed a high
$L/L_\mathrm{Edd}$ accretion rate system -- a NLSy1 at the high mass end
of the normal NLSy1 population. With the current evidence Occam's Razor
favors this explanation over more exotic scenarios as the ejection of the
QSO's black hole in a 3-body interaction or a gravitational recoil event
involving the companion galaxy
\citep[e.g.][]{hoff06,haeh06,merr06,bonn07}. However, these scenarios are
formally not ruled out even if the upper limit can be pushed down by
another $\sim$5 magnitudes. All evidence combined is consistent with a
system of a QSO with ULIRG-size IR emission, residing in an $L^*$ host
galaxy that is in the process of colliding with a substantially more
luminous and possibly more massive companion ULIR-galaxy\footnote{It is
interesting to note that the ``companion'' is close to a factor of 10 more
luminous than the host galaxy. With all uncertainties included it would
still appear as if the typical mass ratio upper limit of 1:3 for the
merging galaxies in a ULIRG system \citep{dasy06} were exceeded
here. However, when using the dynamical masses from \citet{papa08} to
predict a black hole mass in the host galaxy consistent with the
\citet{haer04} relation, we get a merger mass ratio of 1:1 or 1:2.}. Much
deeper high-resolution NIR imaging with a well controlled PSF are the best
way to finally find and trace the here predicted host galaxy (bulge)
component of \he\ co-centered with the QSO nucleus and to estimate its
luminosity and mass directly.

\subsection{Black hole -- galaxy coevolution}
Given the black hole mass and Eddington ratio the accretion rate of the BH
is 1.4~M$_\odot$/yr. At the same time \citet{papa08} derive a star
formation rate from CO of 370~M$_\odot$/yr, predominantly in the companion
galaxy.  Applying a correction factor of 0.5 for mass returned to the
interstellar matter by stellar winds, the stellar mass growth of the whole
\he\ system from star formation is 185~M$_\odot$/yr. The ratio of black
hole accretion and stellar mass growth is then 12/185=6.5\%, which is
substantially higher than the $M_\mathrm{BH}$/$M_\mathrm{bulge}$ relation
for local galaxies of 0.14\% \citep{haer04}.

We can conclude the following: If activity timescales are identical for
star formation and BH accretion, this system grows in black hole mass much
more rapidly than the bulge is required to grow to keep the system on the
$M_\mathrm{BH}$/$M_\mathrm{bulge}$ relation. This is not possible, since
the star formation is taking place in the companion and not the host
galaxy. So in any case a potential maintainance of the relation for this
system, if actually true, needs to be seen as an integral over more than
several 10$^8$ yrs.

On the other hand, a gas consumption timescale of 9.5$\times$10$^7$ yrs --
if we divide the H$_2$ masses and SF rates derived by \citet{papa08} and
account for 50\% mass recycling -- is possibly longer than the luminous
quasar accretion phase. This would add to the requirement, that processes
like the tidal forces of the galaxy interaction redistribute mass, adding
stars to the bulge of the host galaxy. These were to the larger extent
already preexisting in the host galaxies disk or the companion before the
interaction and not created only now. The ``coevolution'' of the host
galaxy and its black hole in \he\ is clearly a two-part process: the
build-up of stellar mass and the build-up of black hole and bulge
mass. The former will take place on timescales of $>$1 Gyr through star
formation, the latter two can ``coevolve'' if seen as an average over
timescales of longer than the BH accretion lifetime, and a few dynamical
timescales for redistribution of stellar orbits of, say, $<$500 Myrs.

\subsection{How many \he s are there?}
\he\ is an unusual object. AGN in ULIRGs are common, but AGN right next to
ULIRGs are not, particularly not luminous QSOs with inconspicuous host
galaxies next to extreme starformers. So is \he\ one of a kind or was it just
the scarceness of IR imaging with 1\arcsec\ resolution and high-resolution CO
maps that prevents us from finding similar objects en masse?

In the higher redshift Universe there was a recent report of a very
similar system \citep[][]{youn08}. LH850.02 at $z=3.3$ is the brightest
submm galaxy in the Lockman hole. Using the Submillimeter Array, the
authors find two components of which one is a ULIRG with intense star
formation, while the other component likely harbors an AGN. At $z>2$
however, objects like this might be quite common, since merging rates and
gas reservoirs were much larger than today. If there existed a substantial
number of similar systems at low redshifts, this would allow to study
mechanisms of the high-redshift Universe at much lower distances.

We try to estimate the frequency of such systems in the local Universe
using the three morphologically best studied samples of quasars at
$0.05\la z<0.43$. We deliberately use optically selected quasars only, as
they have no bias with respect to frequency of merger signatures or
extreme SFRs as IR-selected samples have by construction. In this way
statements about the general population are possible. \citet{jahn04a}
investigated a volume-limited and complete sample of 19 luminous QSOs out
to $z=0.2$. While at least five of these QSOs are seen in intermediate and
late stages of major mergers, only one, HE\,1254--0934, is a likely
ULIRG\footnote{This is a borderline case because it will fall slightly below
or above the ULIRG definition limit depending on if we include upper
limits in 12 and 25$\mu$m or assume the flux to be zero.},
as determined from its IRAS fluxes. It is also among the most distorted
systems, with a companion at $\sim$1\arcsec\ distance from the QSO nucleus.
The companion is more luminous than the host galaxy, and shows a substantial
tidal tail. It looks remarkably similar to \he.

The two other samples are not volume-limited samples, so the selection
function is unclear -- except that these quasars stem from either optical
or radio surveys, but not the IR. \citet{floy04} studied the morphologies
of two intermediate- and high-luminosity samples of ten radio-quiet and
seven radio-loud quasars at $0.29<z<0.43$, using HST-imaging data. Only
one of their 17 quasars shows a distorted geometry similar to \he\
(1237--040 at $z=0.371$) but there exists no information about the total
IR emission or star-formation rates. The IRAS flux limits of 200mJy is
equivalent to upper limits of $L_\mathrm{ir}\sim6\times10^{12}$ $L_\odot$
at $z=0.37$.  ULIRG-strength emission for 1237--040 could have gone
unnoticed by IRAS.

A recent study by \citet{kim08b} determined the morphologies of 45
HST-archived quasars at $z<0.35$. It has one object in common with
\citet{floy04} and three objects with \citet{jahn04a}. Of their sample, three
other objects (HE\,0354--5500, PG\,1613+658, PKS\,2349--01) are clearly
merging with a nearby companion, and are likely ULIRGs as judged from their
IRAS fluxes.  However, only in the case of HE\,0354--5500 the quasar and
companion are still well separated and their envelopes have not yet merged
into a common halo. The two other cases are in a very late merger state and
star-formation will likely occur all over the system.

This adds up to only $\le$3/77 QSOs to possibly be \he-like in the three
samples combined. At $\le 4$\% such systems are indeed rare in the local
Universe. These three quasars however should be investigated in more
detail. It needs to be tested how strong their star-formation actually is,
where in the system it is localized, and if the separated companion is in
any way connected to the AGN-fuelling. If a similar situation as for \he\
is found, the result can set strong constraints on the ULIRG--AGN
evolutionary scenario \citep{sand96} and the creation mechanisms of AGN at
high redshifts. It can contribute to answering the question whether
SF-ULIRG activity in AGN systems is an indicator of a specific mechanism
of AGN fuelling. Or, if these are just the most gas-rich merger-triggered
AGN systems at the top end of SFRs, with a continuous sequence towards
less gas-rich merger-triggered AGN systems. The merging--AGN fuelling
mechanism could be identical from ULIRGs down to the Seyfert regime, where
at some point secular mechanisms become more dominant. Lower SFR systems
could just be the consequence of lower gas mass, but this might only
mildly impact on the -- much smaller -- AGN fuelling rate.

\section{Conclusions}
With new NIR and MIR images to spatially resolve the \he\ system, and in
the light of previously existing data, we find:

\begin{enumerate}

\item The companion galaxy is covered in optically thick and unevenly
 distributed dust. This makes it appear as a collisional ring galaxy in
 the optical, but intrinsically it is smooth and has smooth NIR emission
 increasing towards a pronounced center. The star formation in the
 companion is similar to the strong starburst Arp220, while softer
 M82-like star formation is ruled out. This can reconcile the SFR
 estimates from the optical and FIR. The companion is a star-formation
 powered ULIRG.
 
\item Our MIR image confirms a single warm dust point source at the location
  of the QSO nucleus. This supports a two component dust SED with the warm
  component fully associated with the QSO nucleus, which is an AGN-powered
  ULIRG.

\item A dust-free line of sight to the quasar nucleus is evidence
 that the host galaxy is not obscured by large amounts of dust. However,
 the ULIRG-strength warm IR emission by the nucleus and the upper limit on
 star formation in the host galaxy of substantial 60~M$_\odot$/yr leave
 room for dust.

\item With $H\ge16.9$ the current NICMOS images do not set stronger upper
 limits on the host galaxy of \he. The $V$-band, $H$-band, and
 CO-constraints give $M_V\ge-21.2$ to $M_V\ge-22.0$ depending on the
 assumed dust masses.

\item Flux in the NE-extension of $H=18.8$ is likely associated with the
 QSO's host galaxy. It corresponds to a first lower limit of $M_V<-20.4$
 for the host galaxy. With a black hole of $\sim6.5\pm 2.5
 \times10^7$~M$_\odot$, an accreting rate of 12~M$_\odot$/yr equal to
 super-Eddington accretion, $L/L_\mathrm{Edd}=6.2^{+3.8}_{-1.8}$, the host
 galaxy is consistent with the
 $M_\mathrm{BH}$--$M_\mathrm{bulge}$-relation for normal galaxies. It is
 also consistent with \he\ being a NLSy1 at the high end of the known
 black hole mass distribution. The reason for the high accretion rate is
 unclear but could be connected to \he\ being in an early stage of merging
 with its gas-rich companion.  A more exotic explanation for the system is
 currently not required by any data, but can in the end only be ruled out
 with much deeper, high-resolution NIR images to find the main body and
 bulge of the host galaxy.
 
\item If host galaxy and black hole in \he\ are co-evolving according to
  the local $M_\mathrm{BH}$--$M_\mathrm{bulge}$ relation, it has to occur
  over longer timescales ($\le$500~Myr) and/or the mass growth for the bulge
  is predominantly not caused by the current star formation in the system,
  but by redistribution of preexisting stars.

\item A constellation as in the \he\ system with separate locations of QSO
  nucleus and strongly star forming ULIRG companion might be common at $z>2$
  where gas masses and merger rates were higher, but at a fraction of $\le$4\%
  it is extremely rare in the local Universe.

\end{enumerate}

\acknowledgments
The authors would like to thank E.~F.\ Bell, A.\ Mart\'inez Sansigre, H.\
Dannerbauer, E.\ Schinnerer, K.\ Meisenheimer, F.\ Courbin, P.\ Magain and
H.-R.\ Kl\"ockner for very fruitful discussions and helpful pointers.

Based on observations made with ESO Telescopes at the Paranal Observatory
under programme ID~276.B-5011. Also based on observations made with the
NASA/ESA Hubble Space Telescope, obtained at the Space Telescope Science
Institute, which is operated by the Association of Universities for Research
in Astronomy, Inc., under NASA contract NAS 5-26555. These observations are
associated with program \#10797. This research has made use of the NASA/IPAC
Extragalactic Database (NED).

KJ acknowledges support through the Emmy Noether Programme of the German
Science Foundation (DFG) with grant number JA~1114/3-1. AB is funded by
the Deutsches Zentrum f\"ur Luft- und Raumfahrt (DLR) under grant
50~OR~0404. VC, Research Fellow, thanks Belgian Funds for Scientific
Research. This work was also supported by PRODEX experiment arrangement
90312 (ESA and PPS Science Policy, Belgium).

{\it Facilities:} \facility{ESO VLT (VISIR)}, \facility{HST (NICMOS)}.


\end{document}